# A Generic Cognitive Dimensions Questionnaire to Evaluate the Usability of Security APIs

Chamila Wijayarathna[1], Nalin A.G. Arachchilage[2], and Jill Slay[3]

Australian Centre for Cyber Security
University of New South Wales (UNSW Canberra)
Australian Defence Force Academy, Australia
Email:[1]c.diwelwattagamage@student.unsw.edu.au
[2]nalin.asanka@adfa.edu.au
[3]j.slay@adfa.edu.au

**Abstract.** Programmers use security APIs to embed security into the applications they develop. Security vulnerabilities get introduced into those applications, due to the usability issues that exist in the security APIs. Improving usability of security APIs would contribute to improve the security of applications that programmers develop. However, currently there is no methodology to evaluate the usability of security APIs. In this study, we attempt to improve the Cognitive Dimensions framework based API usability evaluation methodology, to evaluate the usability of security APIs.

## 1 Introduction

In January 2014, hackers posted user-names and telephone numbers of 4.6 million US Snapchat account holders on-line due to an insecure Application Programming Interface (API) used in Snapchat app [12]. Snapchat is one of the most popular mobile apps among teens that allows its users to send and receive "self-destructing" pictures and videos. However, hackers claimed that their intention was to raise public awareness around the insecure API issue and also to put public pressure on Snapchat to get this security flow fixed. As reported in May 2015, Starbucks suffered from a similar fate that hackers drained money from its customers' bank and PayPal accounts through an insecure API [26]. Nevertheless, programmers in the software development industry have been heavily dependent on the use of APIs [34] [33].

An API is a salient part of a reusable software component which acts as the interface where programmers can call the features of the component. Using features of a reusable software component through an API helps the programmer to use them effectively for developing applications, even without a knowledge of implementation details of the component. Therefore, the use of APIs has become an inseparable part of a programmer's life.

One of the functionalities that APIs provide is security. Due to high complexity of security concepts, security related components are designed and implemented by designers specialized in security [16] [33]. Programmers use those components through APIs exposed, which we call as security APIs. Programmers use security APIs to achieve various security functionalities such as authentication and authorization, input validation, encryption, decryption, hashing, etc.

Even though APIs are important in software development process, often they are not very easy to learn and use in software development environment [3] [30] [31] [34]. Less usability of APIs causes to reduce efficiency of programmers where they have to spend significant time to learn the APIs [23]. Also less usable APIs lead programmers to incorrectly use them, which causes unintended behaviors in resulting systems.

The situation is worse with less usable security APIs. When the programmer uses a security API incorrectly, that causes security vulnerabilities in the system s/he develops. In a study Fahl et al. carried out using 13500 popular free Android apps, they found that 8% of the apps are vulnerable to attacks like man in the middle attack, due to improperly using the Secure Socket Layer (SSL)/Transport Layer Security (TLS) APIs [13]. The authors have identified that the cause for this is not only the carelessness of the programmers, but also the usability issues of the SSL/TLS APIs used by programmers for developing those apps.

If the usability of security APIs can be improved, they will be less prone to erroneous usages and therefore, will be less subject to introduce security vulnerabilities to the applications [16] [22]. As per the knowledge of the authors, currently there is no existing methodology to evaluate the usability of security APIs. Thus, in this study our contribution focuses on developing a methodology to evaluate the usability of security APIs.

The rest of the paper is organized as follows. Section 2 presents the related work from existing literature. Section 3 describes the new dimensions and questionnaire we propose in this study. In the final section, we summarize the work presented and conclude with an outlook on future work.

## 2  Related Work

### 2.1  Usability of Security APIs

APIs provide a mechanism for code reuse, where programmers can build their software applications on top of other software components which already exist rather than writing the code from scratch [20] [23] [27] [28] [33] [34]. Hence, effective APIs are important to ensure the better use of the underlying components, and the usability of APIs demands increasing interest [11] [34]. Due to the impact that API usability seems to have, Myers and Stylos [23] suggest that "Following its design, a new API should be evaluated to measure and improve its usability". There has been a number of studies that introduce and use various methods to evaluate the usability of APIs [3] [8] [14] [19] [27] [28] [31]. Some of the most popular methods for evaluating usability of APIs are empirical evaluation [8] [9] [27], heuristic evaluation [19], conducting user studies [3] [29] [30] [31], API peer reviews method [14] and API concepts framework based automated methodology [28].

Even though, number of methods for evaluating API usability have been suggested as mentioned above, evaluating usability of security APIs is still a less attended topic [16] [22] [23]. Security APIs are a subset of APIs which are used to secure the boundary between trusted and untrusted code [1]. Bond [5] defined a security API as "an application programming interface that uses cryptography to enforce a security policy on the interaction between two entities". However, this definition by Bond exclude some APIs which provide security functionalities without using cryptography (eg: input validation APIs such as Open Web Application Security Project(OWASP) esapi). Hence in this study, we consider security APIs as "application programming interfaces that provides developers with security functionalities that enforce one or more security policies on the interaction between atleast two entities", as defined by Gorski and Iacono [16].

Several previous studies have discussed the importance of the usability of security APIs and the effects of the security APIs which are not usable [5] [13] [15] [22] [23] [32] [33]. Fahl et al. [13] and Georgiev et al. [15] list and discuss number of software, which were vulnerable to cyber attacks because of the usability issues in SSL/TLS API that has been used to develop the software. Myers and Stylos [23] also discussed the importance of the usability of security APIs, pointing to the results obtained by Fahl et al. [13]. Weber [32] describes the importance of the usability of APIs such as authentication API provided by Facebook. He suggests that APIs like those should be usable, otherwise cyber-security failures will result on the software which makes use of them.

Wurster and Oorschot [33] highlight that most of the times programmers are not the experts of security, and also most programmers believe that their code is not security critical. Authors suggest that educating all the programmers about security concepts is not feasible and the most feasible solution is making the APIs that they use more secure and usable.

Mindermann [22] discusses the importance of the usability of security APIs for developing more secure software. He claims that the security of developed applications will be far better if the security libraries are more usable. He also highlights the importance of applying usability research for security APIs to deliver more usable security APIs.

Even though the importance of the usability of security APIs has been discussed, only a limited work has been done to achieve this [16] [17]. By referring and analysing the outcomes of existing security studies, Gorski and Iacono [16] list 11 security API specific usability characteristics. According to the authors, this set is not complete, so there can be more characteristics that describe usability of security APIs. Green and Smith [17] also point out 10 rules to create a good crypto API. Furthermore, they urge the need for qualitative and quantitative empirical studies in this area.

From looking at the existing literature, even though different methods have been identified to evaluate the usability of APIs, none of them has been used to evaluate the usability aspects of security APIs. In this study, we try to address this problem and propose a methodology to evaluate the usability of security APIs.

We are proposing an empirical evaluation methodology similar to the one used by Microsoft Visual Studio Usability group in their API usability evaluations [8] [9]. We choose this methodology over other usability evaluation techniques (i.e. heuristic evaluation, API peer review method, API concepts framework and conducting unstructured user studies) due to several reasons. First of all, empirical evaluation requires involvement of programmers who are the actual end users of the API. In our point of view, this is essential for evaluating the usability of security APIs, because security vulnerabilities caused by the usability issues that exist in security APIs, occur when programmers incorrectly use security APIs. Getting them involve in the evaluation process will help evaluators to identify what usability issues presuade programmers to use the API incorrectly. Furthermore, getting end users of the product involved in the usability evaluation process is considered as the gold standard among Human Computer Interaction (HCI) specialists [23] [24]. In addition to that, this methodology requires less expert intervening and also sensitive to wide range of usability aspects compared to API peer review method and API concepts framework. Therefore, we believe that conducting empirical evaluations using Cognitive Dimensions framework will be more effective than using other mentioned methodologies, for evaluating the usability of security APIs.

**2.2  Cognitive Dimensions Framework Based Usability Evaluation**

Cognitive Dimensions of Notation framework was first introduced by Green [18] as a broad-brush discussion tool to discuss usability issues of programming tools. In 1999, Kadoda et al. [21] used this framework to empirically evaluate usability of educational theorm provers. They changed the evaluation procedure by getting end user involved in the evaluation process through a questionnaire. Blackwell and Green [4] acknowledge the importance of this method saying that users do all the work here, so less expert involvement is required. However, this approach used by Kadoda et al. [21] have few drawbacks. Blackwell and Green [4] point out that, since system designer is the person who designs the questionnaire to evaluate the system and selects the dimensions to use, some usability aspects that may important in users perspective will be ignored. Furthermore, Blackwell and Green [4] mention that it adds extra burden since a different questionnaire has to be developed for each system to be evaluated.

As a solution to these problems, Blackwell and Green [4] describe an enhancement for this method which uses a generic questionnaire. They presented a complete questionnaire which covers all 16 cognitive dimensions of the Cognitive Dimensions Notation Framework of Green [18]. There are many advantages of using a generic questionnaire over using a questionnaire specific to a system. When using a generic questionnaire, user do all the work related to the usability evaluation and data retrieved through evaluation will only demonstrate user's judgement. Furthermore, same questionnaire can be used to evaluate any system. Therefore, burden of creating questionnaire per each system has removed here.

In 2004, Clarke [8] presented a methodology used by Microsoft Visual Studio Usability Group to evaluate the usability of APIs. He used the same methodology described by Blackwell and Green [4] with a modified set of cognitive dimensions and a different questionnaire [10]. The framework Clarke used consisted of 12 dimensions which are,

- Abstraction level
- Learning style
- Working framework
- Work-step unit
- Progressive evaluation
- Premature commitment
- Penetrability
- API elaboration
- API viscosity
- Consistency
- Role expressiveness
- Domain correspondence

Clarke alleges that Microsoft Visual Studio Usability Group has proved the relevance and the utility of the cognitive dimensions framework for evaluating API usability, however the usage of the above mentioned questionnaire has not been backed by any empirical evidence. Clarke mentions that he developed this questionnaire based on his experience and the feedbacks of participants who involved in usability tests at Microsoft [10]. Following is the methodology they used for evaluating usability of APIs.

Firstly, experimenters recruit participants and ask them to write code that accomplishes various tasks using the API that need to be evaluated. While participants are doing this, evaluators recorded data such as video records of participants' behaviour and participants' verbal accounts for their actions(Participants were employed in a think-aloud study [2] [6]). After the tasks are completed, the evaluators ask participants to answer the questionnaire [8]. Based on the participants' feedback, evaluators identify the usability issues that exist in the API.

Other researchers have also used this methodology for evaluating the usability of APIs. For an example, Piccioni et al. [27] used the same approach with slight modifications to evaluate the usability of a data persistence library API written in Eiffel. Without using the 12 dimension cognitive dimensions framework introduced by Clarke, they have only considered 4 dimensions which are *understandability*, *abstraction*, *reusability* and *learnability*. They have used their own questionnaire developed based on these dimensions.

As discussed in the previous subsection, we propose the same methodology to evaluate the usability of security APIs. Even though we can use the same methodology described by Clarke [8] to evaluate usability of security APIs, the dimensions and questionnaire he used are not sufficient to do this. This is supported by the fact that Gorski and Iacono [16], and Green and Smith [17] recommend more different characteristics to consider when discussing usability of security APIs. Also, improving usability with respect to some aspects can cause to reduce the security [23]. Thus, when evaluating usability of security APIs, we may have to omit some of the dimensions listed by Clarke and add some new dimensions. Therefore, in this study, we are proposing an enhanced and fine tuned version of Cognitive Dimensions framework and the questionnaire, that can be used to conduct empirical usability evaluations for security APIs.

## 3  Questionnaire Design

We considered Microsoft's version of Cognitive Dimensions framework [8] as the starting point to develop the new questionnaire. Then we improved it by referring to the past studies conducted in this area [16] [17] and by taking usability guidelines those studies have mentioned into consideration.

First, we took 10 rules mentioned by Green and Smith into account. Their first rule is **Easy to learn - even without crypto background**. This is related to the **Learning Style** dimension in the Microsoft's version of Cognitive Dimensions framework. **Learning Style** describes the knowledge about the API that the programmer needs to have before start using the API, and how user would gain the knowledge he requires about the API [7] [8]. However, Green and Smith talk about the cryptographic knowledge requirements that the programmer needs to have. In previous sections, we discussed that most programmers who use security APIs are not security experts. Therefore, **Easy to learn without crypto or security background** is an important aspect to consider when evaluating usability. Since this is related to **Learning Style** dimension, without introducing as a new dimension, we added new questions to Learning Style dimension to cover this aspect.

- Do you think your previous computer security related knowledge made it easy to use the API? What previous knowledge helped in using the API?
- Do you think, if you had previous knowledge of any specific area related to computer security, it would have been easier to use the API? What are those areas you think would have been useful?

We did not consider the **Easy to use - even without documentation**, **Sufficiently powerful to satisfy non-security requirements**, **Hard to circumvent errors - except during testing/development** and **Assist with/ handle end-user interaction** rules. We could not get a proper idea about what the authors tried to convey from these properties by referring to the resources available. Also, since our objective is to improve the Cognitive Dimensions framework to support security APIs, we assume that not using these rules which are not related to security will not reduce the effectiveness of the framework.

**Hard to misuse** is an important rule to consider when evaluating usability of security APIs, because most security related issues occur when programmers misuse security APIs intentionally and unintentionally. This aspect is not covered in the Clarke's version of Cognitive Dimensions framework. Thus, we included this rule with the following questions.

- Have you come up with incidents where you incorrectly used the API and then identified the correct way of doing that? Did the API give any help to identify that you used the API incorrectly? If there were any similar incidents, please explain.
- Did the API give proper error messages in case of exceptions and errors, or did you have to handle them at your programme level? If you had to handle them at your level, please mention the scenario/s.

We decided to omit remaining rules, because Cognitive Dimensions framework already covered those rules by its existing dimensions. **Easy to read and maintain code that uses it** rule says the same as the **Role Expressiveness** dimension. Similarly **Hard to override/change core functionality** rule is covered by **API Eloboration** dimension and **Appropriate to audience** rule is covered by **Learning Style** dimension.

Then we took 11 characteristics of the usability of security APIs suggested by Gorski and Iacono [16] into consideration. The first characteristic they have mentioned is **End-user protection**. This characteristic says that security of an application which uses a security API should not depend on the programmer who develops the application. It could be argued that this is something that needs to be considered when evaluating usability of a security API. Since this aspect is not covered in our questionnaire, we added a new dimension with questions to cover that.

- Do you think the security of the end user of the application you developed, depends on how you completed the task? Or does it depend only on the security API you used?
- If you think security of the end user depends on how you completed the task, in which ways does it depend?

The next characteristic Gorski and Iacono have listed is **Case distinction management**, which refers to the handling of exceptional events and errors that occur related to the API. When these errors and events need to be handled by the programmer, most of the time they do it incorrectly, which leads to vulnerabilities in resulting applications [13] [15] [16]. This is the same issue which we discussed at **Hard to misuse** rule by Green and Smith [17]. Since we added new questions to **Hard to misuse** rule, we didn't add new questions to **Case distinction management** rule again.

We did not include the third characteristic mentioned by Gorski and Iacono, which is **Adherence to security principals**. It says that a security API must follow security guidelines such as "OWASP coding practises" [25]. This is not a property that can be evaluated by observing programmers who use API to implement their applications. So we did not add this characteristic in to the questionnaire.

Fourth characteristic mentioned by Gorski and Iacono is **Testability**, which means API must support reliable test routines written by security experts. This closely relates to **Learning Style** and **Progressive Evaluation** dimensions of the Microsoft's version of Cognitive Dimensions framework. However, those dimensions do not address whether the API provide means to test the security of the application developed using the security API or not. Therefore, we added a new dimension **Testability** with following questions.

- Did you test the security of your application after completing the task using security API? If not, can you explain why?
- If yes, can you explain how did you do that?
- Did the API provide any guidance on how to test the security of the application you developed?

Next characteristic is **Constrainability**, which means letting programmers do configurations related to security causes security vulnerabilities. This is the same we discussed with **API Elaboration** and **Hard to misuse**, so we did not add new questions for **Constrainability**.

**Information obligation** characteristic describes the extent which the API informs the programmer about the security relevant aspects of the security API. Even though the **Penetrability** dimension talks about the API related information exposed by the API to the programmer [7] [8], it does not address whether the API's security related specifics are properly communicated to the programmer. Therefore, without adding a new dimension, we added a new question to the **Penetrability** dimension to cover this characteristic.

- Did the API (including its documentation) provide enough information about the security relevant specifics related to the task you completed? What information was missing or you had to find by referring to external sources?

Next characteristic by Gorski and Iacono, which is **Degree of reliability**, does not talk directly about a property of the API. It talks about the reliability of web resources that the programmer refers while using the API to achieve a task. However, the programmer refers to external unreliable resources (eg: stackoverflow), because API does not expose enough information to the programmer who is using it. We have considered whether the API is providing enough information to the programmer or not, at **Penetrability** dimension. Therefore, we did not add a new dimension to the questionnaire to cover the **Degree of reliability** characteristic.

**Security prerequisites** characteristic says that there are mandatory prerequisites that need to be fulfilled by programmers when using security APIs, which are often unknown and unclear. These prerequisites are also a type of security related information that an API needs to communicate to the programmer, which we described under **Information obligation**. Therefore, we did not add new questions to cover this characteristic, since this is already covered in our questionnaire. Similarly, we did not add new questions to **Execution platform** characteristic. It discusses the target execution platform that API is developed for and whether or not the information has been properly communicated to the programmer. We believe this characteristic is also covered by questions under **Penetrability** dimension.

Next characteristic, which is **Delegation** says that, the security APIs delegating implementation of security functionalities to the application programmer, can cause vulnerabilities in applications that the programmer develops. In **End user protection**, we discussed that security of the developed application should not depend on the application programmer, and we already included questions to cover this. Therefore, we did not included new questions into **Delegation**.

The last characteristic, **Implementation error susceptibility**, says that security API usability research should aim to minimize the error susceptibility. According to the authors, error susceptibility is caused by ignoring the first 10 characteristics that Gorski and Iacono mentioned. Hence, we assumed that this aspect is also covered by previous questions added.

Based on our arguments in this section, we have formed a generic questionnaire by improving Clarke's cognitive dimensions questionnaire to conduct empirical evaluations for security API usability, which contains questions of 15 dimensions(refer Appendix A for the complete questionnaire.

## 4  Conclusion and Future Work

In this work, we improved the version of Cognitive Dimensions framework introduced by Clarke [8] and introduced a generic questionnaire, to evaluate the usability of security APIs. We added new questions into Learning Style and Penetrability dimensions to cover the security related aspects. Furthermore, we introduced 3 new dimensions (i.e **Hard to misuse**, **End-user protection** and **Testability**) with questions which we argued referring to existing literature, to be important for evaluating the usability of security APIs.

As a continuation of this work, we would be conducting empirical studies to prove the validity of the model and the questionnaire proposed.

## Acknowledgements

We would like to thank Steven Clark from Microsoft Visual Studio usability group for providing details about API usability studies at Microsoft. We would also like to thank anonymous reviewers for their feedback.

## References

1. R. Anderson. *Security engineering*. John Wiley & Sons, New York, USA, 2008.
2. N. A. G. Arachchilage, S. Love, and K. Beznosov. Phishing threat avoidance behaviour: An empirical investigation. *Computers in Human Behavior*, 60:185–197, 2016.
3. J. Beaton, S. Y. Jeong, Y. Xie, J. Stylos, and B. A. Myers. Usability challenges for enterprise service-oriented architecture apis. In *2008 IEEE Symposium on Visual Languages and Human-Centric Computing*, pages 193–196. IEEE, 2008.

4. A. F. Blackwell and T. R. Green. A cognitive dimensions questionnaire optimised for users. In *Proceedings of the Twelth Annual Meeting of the Psychology of Programming Interest Group*, pages 137–152, 2000.
5. M. K. Bond. *Understanding Security APIs*. PhD thesis, University of Cambridge, 2004.
6. T. Boren and J. Ramey. Thinking aloud: Reconciling theory and practice. *IEEE transactions on professional communication*, 43(3):261–278, 2000.
7. S. Clarke. Evaluating a new programming language. In *13th Workshop of the Psychology of Programming Interest Group*, pages 275–289, 2001.
8. S. Clarke. Measuring api usability. *Doctor Dobbs Journal*, 29(5):S1–S5, 2004.
9. S. Clarke. Describing and measuring api usability with the cognitive dimensions. In *Cognitive Dimensions of Notations 10th Anniversary Workshop*, page 131. Citeseer, 2005.
10. S. Clarke. Re: [ppigdiscuss] cognitive dimensions questionnaire for evaluating api usability, Oct 2016. Email by: Steven.Clarke@microsoft.com.
11. J. M. Daughtry, U. Farooq, J. Stylos, and B. A. Myers. Api usability: Chi'2009 special interest group meeting. In *CHI'09 Extended Abstracts on Human Factors in Computing Systems*, pages 2771–2774. ACM, 2009.
12. J. Eng. Snapchat hacked, info on 4.6 million users reportedly leaked. http://www.nbcnews.com/business/snapchat-hacked-info-4-6-million- users-reportedly-leaked-2D11833474, 1 January 2014. Accessed: 2016-09-08.
13. S. Fahl, M. Harbach, H. Perl, M. Koetter, and M. Smith. Rethinking ssl development in an appified world. In *Proceedings of the 2013 ACM SIGSAC conference on Computer & communications security*, pages 49–60. ACM, 2013.
14. U. Farooq and D. Zirkler. Api peer reviews: a method for evaluating usability of application programming interfaces. In *Proceedings of the 2010 ACM conference on Computer supported cooperative work*, pages 207–210. ACM, 2010.
15. M. Georgiev, S. Iyengar, S. Jana, R. Anubhai, D. Boneh, and V. Shmatikov. The most dangerous code in the world: validating ssl certificates in non-browser software. In *Proceedings of the 2012 ACM conference on Computer and communications security*, pages 38–49. ACM, 2012.
16. P. L. Gorski and L. L. Iacono. Towards the usability evaluation of security apis. In *Proceedings of the Tenth International Symposium on Human Aspects of Information Security and Assurance*, pages 252–265, 2016.
17. M. Green and M. Smith. Developers are users too: Designing crypto and security apis that busy engineers and sysadmins can use securely. Talk at the USENIX Summit on Hot Topics in Security (HotSec'15).
18. T. R. Green. Cognitive dimensions of notations. *People and computers V*, pages 443–460, 1989.
19. T. Grill, O. Polacek, and M. Tscheligi. Methods towards api usability: A structural analysis of usability problem categories. In *International Conference on Human-Centred Software Engineering*, pages 164–180. Springer, 2012.
20. M. Henning. Api design matters. *Queue*, 5(4):24–36, 2007.
21. G. Kadoda, R. Stone, and D. Diaper. Desirable features of educational theorem provers? a cognitive dimensions viewpoint. In *Proceedings of Psychology of Programming Interest Group conference*, 1999.
22. K. Mindermann. Are easily usable security libraries possible and how should experts work together to create them? In *Proceedings of the 9th International Workshop on Cooperative and Human Aspects of Software Engineering*, pages 62–63. ACM, 2016.

23. B. A. Myers and J. Stylos. Improving api usability. *Communications of the ACM*, 59(6):62–69, 2016.
24. J. Nielsen. *Usability engineering*. Academic Press, Boston, USA, 1994.
25. OWASP. Owasp secure coding practices - quick reference guide. https://www.owasp.org/index.php/OWASP_Secure_Coding_Practices_-_Quick_Reference_Guide, 2010. Accessed: 2016-09-26.
26. J. Pagliery. Hackers are draining bank accounts via the starbucks app. http://money.cnn.com/2015/05/13/technology/hackers-starbucks-app/, 14 May 2015. Accessed: 2016-09-08.
27. M. Piccioni, C. A. Furia, and B. Meyer. An empirical study of api usability. In *2013 ACM/IEEE International Symposium on Empirical Software Engineering and Measurement*, pages 5–14. IEEE, 2013.
28. T. Scheller and E. Kühn. Automated measurement of api usability: The api concepts framework. *Information and Software Technology*, 61:145–162, 2015.
29. J. Stylos and S. Clarke. Usability implications of requiring parameters in objects' constructors. In *Proceedings of the 29th international conference on Software Engineering*, pages 529–539. IEEE Computer Society, 2007.
30. J. Stylos, B. Graf, D. K. Busse, C. Ziegler, R. Ehret, and J. Karstens. A case study of api redesign for improved usability. In *2008 IEEE Symposium on Visual Languages and Human-Centric Computing*, pages 189–192. IEEE, 2008.
31. J. Stylos and B. A. Myers. The implications of method placement on api learnability. In *Proceedings of the 16th ACM SIGSOFT International Symposium on Foundations of software engineering*, pages 105–112. ACM, 2008.
32. S. Weber. Empirical evaluation of api usability and security. https://insights.sei.cmu.edu/sei_blog/2016/01/empirical-evaluation-of-api-usability-and-security.html, 2016. Accessed: 2016-09-08.
33. G. Wurster and P. C. van Oorschot. The developer is the enemy. In *Proceedings of the 2008 workshop on New security paradigms*, pages 89–97. ACM, 2009.
34. M. F. Zibran, F. Z. Eishita, and C. K. Roy. Useful, but usable? factors affecting the usability of apis. In *2011 18th Working Conference on Reverse Engineering*, pages 151–155. IEEE, 2011.

## Appendix

### Appendix A - Complete Questionnaire

**Abstraction Level**

- Do you find the API abstraction level appropriate to the tasks?
- Did you need to adapt the API to meet your needs?
- Do you feel that you had to understand the underlying implementation to be able to use the API?

**Learning style**

- Did you had to learn about different components exposed by API before starting to do anything useful related to your task? What are the components that you had to learn?
- Did you had previous experience working with any of the components of the API? If you had, do you think, that knowledge was essential to do anything useful related to your task?

- Did you had to learn about dependencies between different components exposed by API before starting to do anything useful related to your task? What are the dependencies that you had to learn?
- Did you had to learn about the underlying architecture of the API and other conceptual information before starting to do anything useful related to your task?
- Does the API support learning (the stuff required to complete the task), while you progressing with the task?
- Do you think your previous computer security related knowledge made it easy to use the API? Specifically what previous knowledge helped in using the API?
- Do you think, if you had previous knowledge of any specific area, it would have been easier to use the API? What are they?

**Working Framework**

- What are the information you had to maintain while completing the tasks?
- Which of them were represented in the API you had to use?
- Which of them were not directly represented in API, but was represented in the way that your code is structured?
- Which of them were not represented at all in the API or the code that you were writing?

**Work step unit**

- Does the amount of code required for this scenario seem just about right, too much, or too little? Why?
- Does the amount of code required for each subtask in this scenario seem just about right, too much, or too little? Why?

**Progressive evaluation**

- How easy is it to stop in the middle of the scenario and check the progress of work so far?
- Is it possible to find out how much progress has been made? If not, why not?

**Premature Commitment**

- When you are working with the API, can you work on your programming task in any order you like, or does the system force you to think ahead and make certain decisions first?
- If so, what decisions do you need to make in advance? What sort of problems can this cause in your work

**Penetrability**

- What are the places where you had to distinguish between different methods and classes while you work on your programming task?
- Were you able to find enough information to distinguish between different methods and classes while you work on your programming task? If not, what are the information you think is missing?
- What are the places where you had to understand the context or scope of the particular parts of the API you worked with?
- Were you able to find enough information to understand the context or scope of the particular parts of the API you worked with? If not, what are the information you think is missing?

- What are the places where you had to understand the intricate working details of the API while you work on your programming task?
- Were you able to find enough information to understand the intricate working details of the API while you work on your programming task? If not, what are the information you think is missing?
- Did the API (including its documentation) provide enough information about the security relevant specifics related to the task you completed? What are the information that was missing or you had to find by referring to external sources?

### API Elaboration

- Did you had to extend types exposed by the API by providing their own implementation of custom behavior to accomplish task? What are the types you had to extend? Explain why you needed to extend the original type provided by the API in each case.
- Did you had to replace existing types or introduce new types to accomplish task? What are the types you had to replace/introduce? Explain why you needed to replace existing types or introduce new types in each case.

### API Viscosity

- When you need to make changes to previous work, how easy is it to make the change? Why?
- Are there particular changes that are more difficult to make? Which ones?

### Consistency

- Were there different parts of the API that mean similar things, is the similarity clear from the way they appear? Please give examples.
- Are there places where some things ought to be similar, but the API makes them different? What are they?

### Role Expressiveness

- When reading code that uses the API, is it easy to tell what each section of code does? Why?
- Are there some parts that are particularly difficult to interpret? Which ones?
- When using the API, is it easy to know what classes and methods of the API to use when writing code?

### Domain Correspondence

- Did the types exposed by the API map directly onto the types and concepts you expected? If not, please mention the the types you expected and how it was supported in the API?
- Were there any types exposed by the API do not map directly onto the types and concepts you expected? What are they?

### Hard to Misuse

- Have you came up with incidents where you incorrectly used the API and then identified the correct way of doing that? Did API give any help to identify that you used the API incorrectly? If there any similar incidents, please explain?

- Did the API give proper error messages in a case of exceptions and error, or did you had to handle them at your programme level? If you had to handle them at your level, please mention the scenarios.

**End-user Protection**

- Do you think the security of the end user of the application you developed, depends on how you completed the task? Or does it depend only on the security API you used?
- If you think security of the end user depends on how you completed the task, in which ways does it depend?

**Testability**

- Did you tested the security of your application after completing the task using security API? If not, why?
- If yes, how did you do that?
- Did the API provided any guidance on how to test your application?